\documentclass[aps,prl,twocolumn,showpacs,floatfix]{revtex4-1}

\usepackage{amsmath}
\usepackage{amssymb}
\usepackage{hyperref}
\usepackage[pdftex]{graphicx}
\usepackage{color}
\usepackage{bm}

\renewcommand{\d}{\mathrm{d}}

\newcommand{\bb}[1] {\breve{ #1}}

\newcommand{\re}{\mathrm{Re}}

\newcommand{\dd}[1]{\mathrm{d}^n\bb{#1}\,}
\renewcommand{\imath}{\mathrm{i}}

\begin{document}

\bibliographystyle{apsrev}

\title{New equation for lagrangian incompressible fluid flows applied to turbulence}

\author{\surname{Olivier} Poujade}

\affiliation{CEA, DAM, DIF, F-91297 Arpajon, France}

\date{\today}

\begin{abstract}

Theoretical developments in the field of Lagrangian turbulence are made difficult by the fact that equations governing the evolution of lagrangian flows are implicit contrary to eulerian flows. In this article, an {\it exact} explicit equation for incompressible lagrangian fluid flows at high-Reynolds number is constructed. The method to arrive at the equation of motion and the proof that it describes the motion of an incompressible fluid are provided. A truncated version of this new equation is used to show how the lagrangian turbulent spectrum ($E_\mathrm{lag}(\omega)$) could be inferred. This exercise showed a complex interrelation between the stirring force field and the flow itself in the lagrangian turbulence framework whereas the stirring is not affected by the flow in the eulerian point of view. The result is that $E_\mathrm{lag}(\omega)\propto \varepsilon\,\omega^{-2}$ seems independent upon the way the fluid is stirred in the inertial range for a given dissipated power $\varepsilon$. It also showed that $E_\mathrm{lag}(\omega)\sim \omega^{-s}$ with $0\leq s\leq 1/2$ (depending on the stirring) at low-$\omega$ and $\sim\omega^{-4}$ in the viscous range at high-$\omega$.

\end{abstract}

\maketitle

Kolmogorov \cite{kolmo,kolmo62} and Obukhov \cite{obuk} predicted the eulerian spectrum for homogeneous and isotropic turbulence, $E_\mathrm{eul}(k)\propto \varepsilon^{2/3}\,k^{-5/3}$, from two insightful hypothesis of similarity and elementary dimensional analysis. The first serious attempts to understand this result, also considered by many as the birth of modern statistical theory of turbulence, is due to Kraichnan \cite{kraich1} with its direct-interaction-approximation, along with lagrangian improvement \cite{kraich2}, and Wyld's \cite{wyld} field theoretic techniques at the turn of the sixties. Since then, theoretical eulerian turbulence has developed through the years using renormalized perturbation theory \cite{mccomb}, renormalization groups \cite{yakhot}, lagrangian renormalized approximation \cite{kaneda1, kaneda2}, diagrammatic techniques \cite{lvov},  and other mathematical tools borrowed from statistical or quantum field theory such as instanton \cite{instanton}, operator product expansion \cite{falko3}, Feynman's path integrals \cite{polya} and more.

\par 

Along the same line of reasoning, the lagrangian turbulent spectrum was predicted to be $E_\mathrm{lag}(\omega)\propto \varepsilon \,\omega^{-2}$ by Kolmogorov \cite{kolmo,kolmo62}. Much less theoretical developments have been devoted to its understanding \cite{kraich70, falko, falko2} whereas, at the opposite, experimental techniques, such as optical tracer-particle tracking \cite{sato} or ultrasonic Doppler tracking \cite{mord}, have strongly improved. Direct numerical simulations, also, have progressed \cite{popesim, bif1, falk3} rapidly during the last decade. Reasons why the theoretical understanding of eulerian turbulence is more advanced than lagrangian turbulence are, in a large part, to be found in the complexity of the equations involved. The eulerian equation of incompressible viscous fluid motion (Navier-Stokes equation) is a non-linear integro-differential equation which is explicit with respect to the velocity field $\bm{v}(\bm{x},t)$. In contrast, the lagrangian velocity field $\partial_t{\bm{X}}(\bm{x}, t)$, where $\bm{X}(\bm{x}, t)$ is the position of the fluid particle at time $t$ that originates from $\bm{x}$ at $t=0$, can either be found as a solution of the implicit equation \mbox{$\partial_t{\bm{X}}(\bm{x}, t)=\bm{v}(\bm{X}(\bm{x}, t), t)$}, which requires the knowledge of the eulerian velocity field $\bm{v}$, or as a solution of an even more non-linear integro-differential equation \cite{lagfd}. 

\par 

It is the purpose of the first part of this paper to derive an equation of lagrangian incompressible fluid motion that is explicit in the field. The method relies on a least action principle using a penalization technique that ensure the incompressibility condition and the explicitness of the equation. In the second part of this paper, the lagrangian turbulent spectrum is analytically derived from a simplified version of this new equation of lagrangian fluid flow to illuminate its predictive potential.

\vspace{.2cm}

In the eulerian viewpoint, $\partial_i v_i=0$ only states that the rate of change of the volume of fluid particles is zero. The lagrangian framework is more flexible. Indeed, either (i) the rate of change of the volume of fluid particles is zero, since $\partial_i v_i=\partial_t\left(\ln\left[\det(\bm{\partial X})\right]\right)=0$, or (ii) the volume of fluid particles is constant, since $\det(\bm{\partial X})=1$ (initially $\partial_i X_j(\bm{x},0)=\delta_{ij}$). That last constraint can be integrated into a least action Lagrangian density 
\begin{align}
{\cal L}[\xi]=(\partial_t \xi_i)^2/2-c^2(\det_\mathrm{n}(\bm{\delta+\partial\xi})-1)^2 \label{lagg}
\end{align} where $\bm{X}$ is replaced by the deviation field $\bm{\xi}$ such that $\bm{X}(\bm{x},t)=\bm{x}+\bm{\xi}(\bm{x},t)$. The first term in the rhs of (\ref{lagg}) is kinetic energy. The second term is constraint (ii) incorporated using the penalty method (encountered in applied mathematics \cite{ghid} or as a gauge fixing mechanism in quantum field theory \cite{qft} for instance). It involves a constant $c$, with dimension of velocity that needs to be pushed to infinity to enforce the incompressibility constraint. The integer $n$ is the dimensionality of space. The minus sign in front of $c^2$ in the Lagrangian density was chosen so that the derived Hamiltonian density,
 \begin{align}
{\cal H}[\xi]=(\partial_t \xi_i)^2/2+c^2(\det_\mathrm{n}(\bm{\delta+\partial\xi})-1)^2,\label{ham}
\end{align} is identically positive. Its space integral over the entire domain is a constant as time goes by thanks to Noether's theorem (invariance of the formal expression of ${\cal L}$ with respect to time) and then
\begin{align}
\exists\,K(c)\in\,\mathbb{R}^+, ~\forall\, t\in\, \mathbb{R},\,\,\int_D\,{\cal H}(\bm{x}, t, c)\,\mathrm{d}^n\bm{x}=K(c),\label{defH}
\end{align} where $K(c)$ is independent of time $t$. Since $\cal{H}$ is positive, that means $\forall c\in\mathbb{R}$, $0\leq\int_D\,\left[\det_\mathrm{n}(\bm{\delta+\partial\xi})-1\right]^2\,\mathrm{d}^n\bm{x}\leq K(c)/c^2$. One would like to show that when one takes the limit $c=+\infty$, one enforces the incompressibility condition $ \det_\mathrm{n}(\bm{\delta+\partial\xi})=1$ everywhere and at all time. To do so, one need to show that $K(c)/c^2$ goes to 0 as $c$ approaches $+\infty$. 

\par 

The most straightforward way is to calculate $K(c)$ using (\ref{defH}) at $t=0$ since it holds for all time. The closed form of the determinant in eq.(\ref{lagg}) is needed and is provided here for space dimension $n=$3:
\begin{align}
&\det_\mathrm{n=3}(\bm{\delta+\partial\xi})-1=\partial_i \xi_i+(\partial_i \xi_i\partial_j \xi_j-\partial_i \xi_j\partial_j \xi_i)/2\nonumber\\
&+(\frac{\partial_i \xi_i\partial_j \xi_j\partial_k \xi_k}{6}-\frac{\partial_i \xi_i\,\partial_j \xi_k\partial_k \xi_j}{2}+\frac{\partial_i \xi_j\,\partial_j \xi_k\partial_k \xi_i}{3}), \label{nlin3}
\end{align} with linear, quadratic and cubic terms. The expression for $n=$2 is made of the linear and quadratic terms and for $n=$1 it is made of the linear term only (it is for this formal property that we use the deviation field $\bm{\xi}$ instead of the coordinate field $\bm{X}$). Therefore, the free Lagrangian, quadratic in the field (that will produce the linear part of the equation of motion), is ${\cal L}_\mathrm{free}=(\partial_t \xi_i)^2/2-c^2(\partial_i\xi_i)^2$, and does not depend upon space dimension $n$. The interacting Lagrangian, ${\cal L}_\mathrm{int}={\cal L}-{\cal L}_\mathrm{free}$ ($=0$ if $n=$1), will produce the non-linear terms in the equation of motion and does depend upon $n$. The solution of the equation of motion that extremizes the action ${\cal S}[\bm{\xi}]=\int\,{\cal L}[\bm{\xi}]\,\mathrm{d}^nx\,\mathrm{d}t$ verifies $\frac{\delta{\cal S}[\bm{\xi}]}{\delta\xi_i}=0$, that is to say
\begin{align}
-\partial_t^2\xi_i+c^2\,\partial_i(\partial_j\xi_j)+\frac{\delta{\cal L}_\mathrm{int}}{\delta\xi_i}=0,\label{eqnlin}
\end{align} where $\delta{\cal L}_\mathrm{int}/\delta\xi_i$ is a sum of non-linear terms of the form $c^2\,\partial(\partial\xi\cdots\partial\xi)$ up to third order if $n=$2 and up to fifth order if $n=$3 (by inspection of eq. (\ref{nlin3})). 

\par

In order to constraint the initial conditions for $c<+\infty$ it should first be noticed that the limit initial conditions are $\xi_i(\bm{x}, t=0, c=+\infty)=0$, $\partial_j\xi_i(\bm{x}, t=0, c=+\infty)=0$ and $\partial_t\xi_i(\bm{x}, t=0, c=+\infty)=v_i(\bm{x})$. Indeed, this means that as $c$ approaches $+\infty$, the $\xi$'s and their space derivatives should tend to zero. Then, for $c$ large enough and around $t=0$, one can conclude that the non-linear terms in the equation of motion are negligible and that the equation of motion is reduced to $\partial_t^2\xi_i=c^2\,\partial_i(\partial_j\xi_j)$ which should, in turn, be equal to $-\partial_i p$, where $p$ is the initial pressure field, in order to be compatible with the Euler equation at $t=0$. If we were to take the divergence $\partial_i\xi_i(\bm{x}, t=0, c)=0$ for all $c$ we would get $\partial_t^2\xi_i=d_t v_i=0$ instead of $\partial_t^2\xi_i=d_t v_i=-\partial_i p$. This is the reason why this divergence should in fact be
\begin{align}
\partial_i\xi_i(\bm{x}, t=0, c)=-p(\bm{x})/c^2+o(1/c^2).\label{condinit}
\end{align} This is in agreement with the limit initial conditions and enough to calculate $K(c)\!=\!\int\,{\cal H}\,\mathrm{d}^n\bm{x}$ from (\ref{defH}). 

\par 

Since $K(c)\!=\!\int\,(\partial_t\xi_i)^2/2 \,\mathrm{d}^n\bm{x}\!+\!\int\,c^2(\partial_i\xi_i)^2\,\mathrm{d}^n\bm{x}$ at early time (non-linear terms are negligible), its value can be deduced straightforwardly. If one calls $u_\mathrm{rms}^2\!=\frac{1}{V}\!\int_{\cal D}\,v_i^2(\bm{x})\,\mathrm{d}^n\bm{x}$ (the initial kinetic energy of the flow) and $p_\mathrm{rms}^2\!=\frac{1}{V}\!\int_{\cal D}\,p^2(\bm{x})\,\mathrm{d}^n\bm{x}$, which are two finite numbers, one can deduce that $K(c)/c^2=V\,(u_\mathrm{rms}^2/2 c^2+p_\mathrm{rms}^2/c^4)$ which goes to zero as $c$ approaches $+\infty$ as anticipated. Therefore, provided the correct initial conditions, \mbox{$\xi_i(\bm{x}, t=0, c)=-\partial_i\Delta^{-1}p(\bm{x})/c^2$}, the incompressible constraint is enforced as c goes to infinity. This is the reason why, equation (\ref{eqnlin}), in the limit $c=+\infty$, is as good an equation as Euler equation to describe a free incompressible inviscid flow (from the lagrangian point of view).

\par

Now, in order to study turbulence, two crucial ingredients need to be added: viscosity and stirring. The actual newtonian viscosity term, although linear in the eulerian picture ($\nu\,\Delta v_i$), is strongly non linear in the lagrangian framework. This is not an issue when concerned with high Reynolds flows since the precise mechanism of dissipation is not important (since $\nu\rightarrow 0$). It is the existence of such mechanism ($\nu\neq 0$) that is crucial to the phenomenology of turbulence (time-reversal symmetry breaking). For simplicity reasons, a term with the same formal expression ($\nu\,\Delta\partial_t\xi_i$) is included in equation (\ref{eqnlin}). It is equivalent to the NS term for small deviation ($\bm{\xi}$) and will dissipate the injected energy at small scales as will be described later.  The incompressibility constraint is still enforced in the limit $c=+\infty$ since the viscous term invariably decreases $K(c)$ defined in (\ref{defH}). 

\par

A forcing field, $f_i(\bm{x},t)$, should also be added to equation (\ref{eqnlin}) to model the effect of an outside operator stirring the fluid. This forcing is assumed to be statistically homogeneous, isotropic and stationary. The statistic is assumed to be Gaussian which, to be complete, only requires the knowledge of the second moment \mbox{$\langle f_i(\bb{k}_1)\,f_j(\bb{k}_2)\rangle=F_\mathrm{lag}(\bb{k}_1)\,P_{ij}(\bm{k}_1)\,\delta(\bb{k}_1+\bb{k}_2)$}, where \mbox{$\bb{k}=(\bm{k},\omega)$} and \mbox{$\delta(\bb{k}_1+\bb{k}_2)=\delta(\omega_1+\omega_2)\,\delta^n(\bm{k}_1+\bm{k}_2)$} to shorten notations. The brackets $\langle . \rangle$ stand for the ensemble average of various realizations of $f_i$. 

The exact form of $F_\mathrm{lag}$ is not a simple question. Indeed, in the eulerian framework, $F_\mathrm{eul}$ is not part of the problem of turbulence for it is imposed by the stirring operator in the eulerian (laboratory) coordinate system and one is interested in how the fluid reacts to that solicitation. On the contrary, the form of $F_\mathrm{lag}$ in the lagrangian point of view depends upon the eulerian stirring (imposed by the outside operator) but, above all, does depend on the flow itself (since the force field is transported by the flow) and, in this way, is part of the turbulent problem. But it just so happens that the turbulent flow itself does depend on the stirring. This entangled relation between the flow and the lagrangian stirring will be instrumental in getting the lagrangian turbulent spectrum into closed form. 

\par 

The resulting equation of a lagrangian forced incompressible, high Reynolds fluid flow is given by 
\begin{align}
\partial_t^2\xi_i=c^2\partial_i(\partial_\ell\xi_\ell)+\nu\,\Delta\partial_t\xi_i+\frac{\delta {\cal L}_\mathrm{int}}{\delta \xi_i}+f_i,\label{plm}
\end{align} with $c\rightarrow\infty$ and $\nu\rightarrow 0$, and is explicit in the field $\bm{\xi}$.

\par 

In the second part of this article, the consequences of a truncated version of (\ref{plm}), where the interaction term is cast away, is investigated in the realm of lagrangian turbulence. The equation so obtained, although linear in the deviation field,
\begin{align}
\partial_t^2\xi_i=c^2\partial_i(\partial_\ell\xi_\ell)+\nu\,\Delta\partial_t\xi_i+f_i,\label{eqsimp}
\end{align} enables to grasp the potential of the full equation (\ref{plm}), as complex as it may seem. It must be noticed that the linear term $\partial_t^2\xi_i$ in the lagrangian view point is equivalent to the non linear term $\partial_t v_i+v_j\,\partial_j v_i$ in the eulerian framework. Therefore, equation (\ref{eqsimp}) contains all the fundamental ingredient for turbulence: non linearity, non locality, vanishingly small but non zero viscosity and stirring. 

\par 

In what follows, we do not derive the turbulent spectrum of (\ref{plm}) but rather the turbulent spectrum of (\ref{eqsimp}). In diagrammatic terms \cite{lvov}, it amounts to saying that we are searching for the statistical properties of a turbulent flow at tree level. This is already enough to be in qualitatively  good agreement with experiments and numerical simulations. 

\par 

The Green function, or propagator, of (\ref{eqsimp}) can easily be solved in the Fourier space. It is defined in such a way that $\xi_i(\bb{k})=G_{ij}(\bb{k})\,f_j(\bb{k})$ and is solution of \mbox{$-(\omega^2+\imath\,\nu\,k^2\,\omega)\,G_{ij}+c^2\,k_i\,k_\ell \,G_{\ell j}=\delta_{ij}$}. In closed form, it can be expanded in powers of $1/c^2$ as
\begin{align}
G_{ij}(\bb{k})&=-\frac{P_{ij}(\bm{k})}{\omega\,(\omega -\imath\,\nu\,k^2)}+\nonumber\\
&\frac{k_i\,k_j}{k^2}\,\sum_{p\geq 0}\,\frac{1}{c^{2(p+1)}}\,\frac{\omega^p\,(\omega -\imath\,\nu\,k^2)^p}{k^{2(p+1)}},\label{eqgij}
\end{align} where $P_{ij}(\bm{k})=\delta_{ij}-k_i k_j/k^2$ is the projection tensor. The finite part, that does not depend on $c$, is solenoidal (in factor of $P_{ij}$) and the longitudinal component, in factor of $k_i\,k_j$, is of order $1/c^2$ (and smaller). This last part is unimportant at the diagrammatic tree level (that we will be considering in this article) but it contributes to higher order loop diagrams when non-linearities are taken into account (in future developments). 

\par

As long as the flow is statistically homogeneous and stationary, any physical quantity $A$ and $B$ verify $\langle A(\bb{k}_1)\,B(\bb{k}_2)\rangle=F_{AB}(\bb{k}_1)\,\delta(\bb{k}_1+\bb{k}_2)$ and then \mbox{$\langle A(\bm{x},t)\,B(\bm{x},t+\tau)\rangle=\int\,\dd{k} F_{AB}(\bb{k})\,e^{i\,\omega \,\tau}$} where \mbox{$\dd{k}=\d^n\bm{k}\,\d\omega$}. The lagrangian turbulent energy spectrum is defined by $F_{\bm{u}\bm{u}}=E_\mathrm{lag}$ where \mbox{$\bm{u}=\bm{v}(\bm{X}(\bm{x},t))$} is the lagrangian velocity field. The lagrangian power injected spectrum is defined by $F_{\bm{u}\bm{f}}=\Pi$ where $\bm{f}$ is the lagrangian stirring force field. They verify
\begin{align}
\langle \dot{\xi}_i(\bb{k}_1)\,\dot{\xi}_j(\bb{k}_2)\rangle=E_\mathrm{lag}(k_1,\omega_1)\,P_{ij}(\bm{k}_1)\,\delta(\bb{k}_1+\bb{k}_2),\label{eqE}\\
\re[\langle \dot{\xi}_i(\bb{k}_1)\,f_j(\bb{k}_2)\rangle]=\Pi(k_1,\omega_1)\,P_{ij}(\bm{k}_1)\,\delta(\bb{k}_1+\bb{k}_2).\label{eqPi}
\end{align} The mean square value of the flow velocity $u^2_\mathrm{rms}$ and the power injected in the flow (that equals the dissipation $\varepsilon$ for \mbox{$n$ $>$ 2} in the stationary regime) are measurable quantities and are respectively \mbox{$u^2_\mathrm{rms}=\int\,k^{n-1}\,E_\mathrm{lag}(k,\omega)\,\mathrm{d}k\,\mathrm{d}\omega$} and \mbox{$\varepsilon=\int\,k^{n-1}\,\Pi(k,\omega)\,\mathrm{d}k\,\mathrm{d}\omega$}. 

\par

Both spectrum can be calculated using the fact that the Fourier transform of the displacement, in eq. (\ref{eqgij}), when $c$ goes to $+\infty$, is simply
\begin{align}
\xi_i(\bb{k})=\frac{P_{i\ell}(\bm{k})}{\omega\,(\omega-\imath\,\nu\,k^2)}\,f_{\ell}(\bb{k}).
\end{align} This result can then be injected in (\ref{eqE}) and (\ref{eqPi}) and it is found that
\begin{align}
E_\mathrm{lag}(k,\omega)&=\frac{1}{\omega^2+\nu^2\,k^4}\,F_\mathrm{lag}(k,\omega),\\
\Pi(k,\omega)&=\frac{\nu\,k^2}{\omega^2+\nu^2\,k^4}\,F_\mathrm{lag}(k,\omega).
\end{align} 

In order to generate developed turbulence, it is customary to inject mechanical power from a space-correlated and time-uncorrelated stirring force field at a given wave number $k_\mathrm{inj}$ in the eulerian coordinate system: $F_\mathrm{eul}(k,\omega)=\delta(k-k_\mathrm{inj})\,\delta(\omega)$.

In the lagrangian coordinate system, space and time-correlation need to be discussed. The time correlation is assumed to be exponentially decreasing, $e^{-\mid t\mid/\tau}$ which turns into $1/(\omega^2+\Omega^2)$ in Fourier space, with a time constant $\tau=2\pi/\Omega$ to be determined later by the theory (uncorrelated would correspond to $\tau=0$). The formal expression of space correlation in lagrangian coordinate should not depend on its formal expression in eulerian coordinate after few turn-over times. Indeed, any $F_\mathrm{eul}(k)$ provides possible realizations of the force field $f_i^\mathrm{eul}(\bm{x},t)$ in the eulerian coordinate. At a given time $t$, the corresponding lagrangian force field is $f_i^\mathrm{lag}(\bm{x},t)=f_i^\mathrm{eul}(\bm{X}(\bm{x},t),t)$. It is the exact same function where space coordinates have been randomly scrambled by the flow. The density spectrum of such a field is a white noise. But, because $\langle f^2\rangle=\int F_\mathrm{eul}(k,\omega) \mathrm{d}\omega \mathrm{d}^3 k=\int F_\mathrm{lag}(k,\omega) \mathrm{d}\omega \mathrm{d}^3 k$ is finite, so must be the extension of the white noise in Fourier space. That is the reason why, the space power spectrum of the lagrangian stirring force field is $\Theta(\eta_f-k)$ where $\Theta$ is the Heaviside function ($\Theta(x)=1$ if $x>0$  and 0 otherwise) and $\eta_f$ is a cut-off wave number that will also be determined later by the theory. In a nutshell,
\begin{align}
 F_\mathrm{lag}(k,\omega)=F_0\,\frac{k^\ell}{\omega^2+\Omega^2}\,\Theta(k-\eta_f),\label{fstir}
\end{align} where we have assumed a general power law dependence ($k^\ell$) up to a certain scale $\eta_f$ in order to show that the lagrangian turbulent spectrum in the inertial range is independent of $\ell$ even though it is expected that $\ell=0$ (from the discussion above), 

\par 

In the end, 
\begin{align}
E_\mathrm{lag}(k,\omega)&=\frac{F_0\,k^\ell\,\Theta(k-\eta_f)}{(\omega^2+\nu^2\,k^4)(\omega^2+\Omega^2)},\label{ee1}\\
\Pi(k,\omega)&=\frac{F_0\,\nu\,k^{2+\ell}\,\Theta(k-\eta_f)}{(\omega^2+\nu^2\,k^4)(\omega^2+\Omega^2)}.
\end{align} The exact integration over $\omega$ gives 
\begin{align}
E(k)&=\frac{\pi\,F_0\,k^{\ell-2}\,\Theta(k-\eta_f)}{\Omega^2\,\nu+\Omega\,\nu^2\,k^2},\\
\Pi(k)&=\frac{\pi\,F_0\,\nu\,k^\ell\,\Theta(k-\eta_f)}{\Omega^2\,\nu+\Omega\,\nu^2\,k^2}.
\end{align}

Now for the integration over $k$ we need to distinguish the dimensionality of space. The measure of integration in $n$ dimension being $\gamma(n)\,k^{n-1}\,\mathrm{d}k$, the result is $u_\mathrm{rms}^2=\frac{\pi\,F_0}{\Omega\,\nu^2}\int_{\eta_L}^{\eta_f} \frac{k^{\ell+n-3}\,\mathrm{d}k}{\Omega/\nu+k^2}$ and $\varepsilon=\frac{\pi\,F_0}{\Omega\,\nu}\int_{\eta_L}^{\eta_f} \frac{k^{\ell+n-1}\,\mathrm{d}k}{\Omega/\nu+k^2}$ which gives
\begin{align}
\frac{u_\mathrm{rms}^2}{\pi\,F_0}&=\frac{\Omega^{\frac{\ell+n-6}{2}}}{\nu^{\frac{\ell+n}{2}}}\,g_{\frac{\ell+n-4}{2}}\left(\frac{\nu\eta_f^2}{\Omega}\right),\label{u1}\\
\frac{\varepsilon}{\pi\,F_0}&=\frac{\Omega^{\frac{\ell+n-4}{2}}}{\nu^{\frac{\ell+n}{2}}}\,g_{\frac{\ell+n-2}{2}}\left(\frac{\nu\eta_f^2}{\Omega}\right),\label{u2}
\end{align} as long as $n$+$\ell$ $>$ 2, where $g_p(x)=\frac{1}{2}\int \frac{x^p\,\mathrm{d}x}{1+x}$. For $n$=3, $u_\mathrm{rms}^2$ and $\varepsilon$ do not depend upon large scales ($\eta_L^{-1}$) because $\nu\eta_L^2$ can be arbitrarily small since $\eta_L$ is fixed and $\nu$ is vanishingly small (but non zero). It does not hold in two dimensions ($n$=2) for $\ell$=0 because $g_{-1}$ behaves logarithmically and the lower bound $\nu\eta_L^2/\Omega$ can no longer be cast away as $\nu$ approaches zero. 
\par 
The ratio of eqs.(\ref{u1}) and (\ref{u2}), where $F_0$ cancels out, allows for the determination of $\eta_f$. As long as $n$+$\ell$ $>$2, if $\varepsilon$, $u_\mathrm{rms}$ and $\Omega$ are fixed, it is clear from the resulting equation that $\nu\eta_f^2=\omega_L$ is a constant which means that $\eta_f$ ($\propto \nu^{-1/2}$) increases indefinitely as $\nu$ approaches zero. This is why, if $\Omega$ is large enough ($\Omega\gg u_\mathrm{rms}^2/\varepsilon$), the $g$-functions, in eqs.(\ref{u1}) and (\ref{u2}), can be approximated by their Taylor expansion around $x$=0, that is to say $g_p(x)\approx x^{p+1}/2(p+1)$ and it is found that 
\begin{align}
 \omega_L&=\left(\frac{\ell+n}{\ell+n-2}\right)\,\frac{\varepsilon}{u_\mathrm{rms}^2}\\
 F_0&=\frac{(\ell+n-2)^{(\ell+n)/2}}{\pi(\ell+n)^{(\ell+n-2)/2}}\frac{(\nu\,u_\mathrm{rms}^2)^{(\ell+n)/2}\,\Omega^2}{\varepsilon^{(\ell+n-2)/2}}.\label{fzero}
\end{align}
\par
\begin{figure}[h] 
\includegraphics[width=8cm]{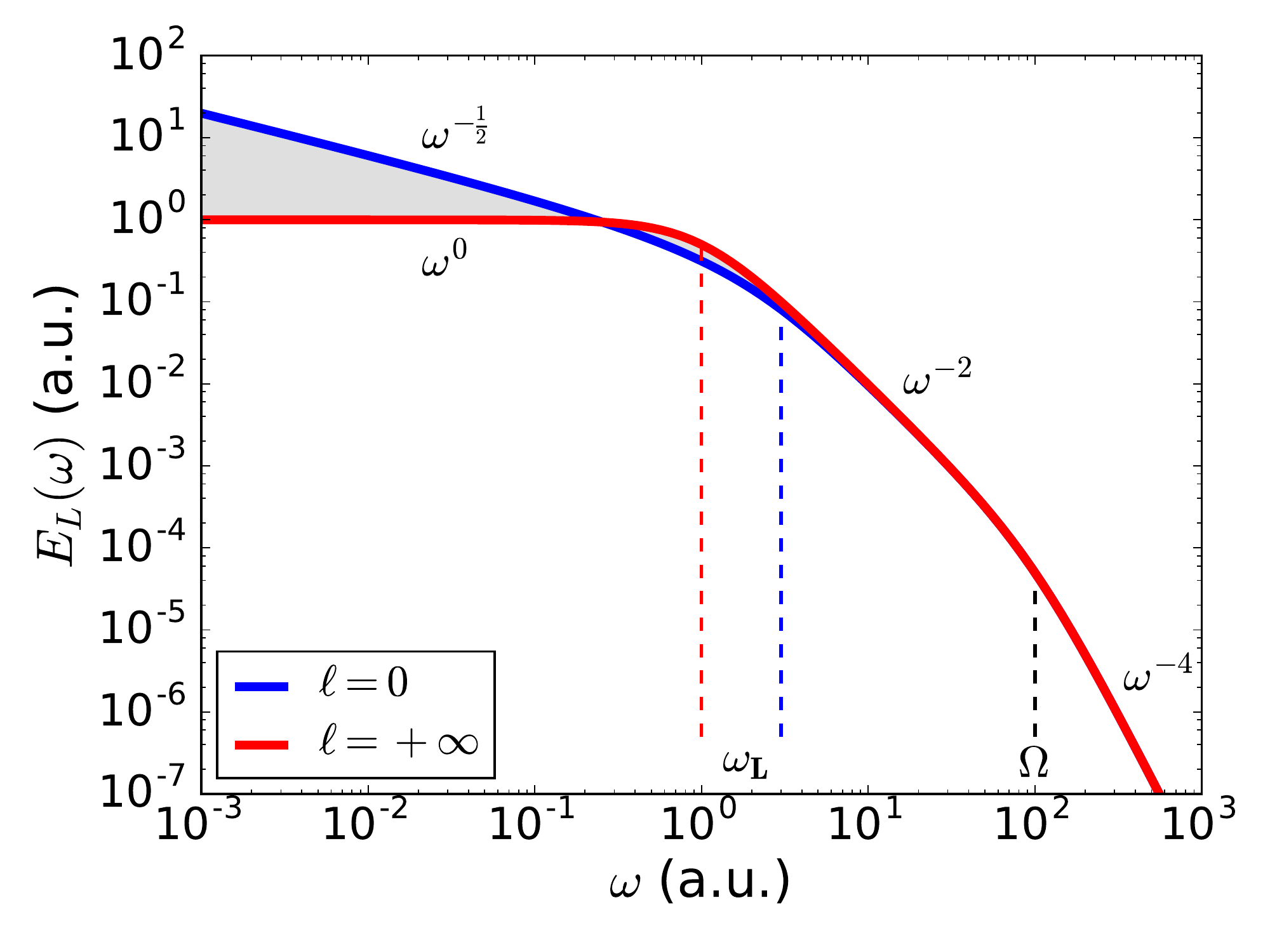}
\caption{In three dimensions, $n=3$, the blue curve is the expected spectrum $E_\mathrm{lag}(\omega)$, eq.(\ref{ee2}), for $\ell=0$. The gray shaded area corresponds to the variability of the spectrum as $\ell$ is varied up to $+\infty$ where the spectrum turns into the red curve.}\label{fig:spec}
\end{figure}

 
In order to calculate $E_\mathrm{lag}(\omega)$, eq.(\ref{ee1}) needs to be integrated over $k$. If $n$+$\ell$ $>$2, it does not depend upon large scales ($\eta_L$) as $\nu$ approaches zero. In this limit,
\begin{align}
 E_\mathrm{lag}(\omega)=\frac{2}{\pi}\,\varepsilon\,\omega^{-2}\,\frac{\Omega^2}{(\omega^2+\Omega^2)}\,h_{\frac{\ell+n}{2}}\left(\frac{\omega_L}{\omega}\right),\label{ee2}
\end{align} where $h_p(x)=p\, x^{-p}\,\int_0^x \frac{y^{p-1}\,\mathrm{d}y}{1+y^2}$. The lagrangian spectrum is plotted in Fig.(\ref{fig:spec}) for the two extreme values $\ell=0$ (which is expected) and $\ell=+\infty$. When integrated over $\omega$ (from 0 to $+\infty$; this is important in order to reproduce the factor 2 in front) it gives precisely $u_\mathrm{rms}^2$. It is to be noticed that the spectrum does not depend upon $\nu$. This is important from a theoretical standpoint because the viscosity $\nu$, used in our development, is not the physical Newtonian viscosity but was tailored to mimic the viscous anomaly (high Reynolds NS flows, in the limit $\nu\rightarrow 0$, do not behave as Euler flows where $\nu=0$) and was meant to be brought to zero at the end of calculations.

\par 

When $u_\mathrm{rms}^2/\varepsilon\ll\omega\ll\Omega$, the h-function in eq.(\ref{ee2}) can be Taylor expanded ($h_p(x)\approx 1$ when $x\ll 1$) and the resulting lagrangian turbulent kinetic energy is
\begin{align}
E_\mathrm{lag}(\omega)=C^\mathrm{theo}_K\,\varepsilon\,\omega^{-2}, \label{reslag}
\end{align} which is the lagrangian turbulent spectrum derived in the literature from dimensional analysis. The analytical derivation provided here produced $C^\mathrm{theo}_K=2/\pi$ which is approximately 0.64. It is important to notice that eq. (\ref{reslag}) is not only independent of $\nu$ but it is also independent of the dimensionality $n$ and of the stirring spectrum power $\ell$ as long as $n$+$\ell$ $>$2. Since $\ell=0$ is expected (lagrangian stirring), it means that this result is valid in three dimensions but not in two dimensions. This is not surprising because a large scale dissipation \cite{alphau}, such as a linear friction term ($-\alpha\,\bm{v}$), must be added to the NS equation in order to sink the inverse energy cascade when $n=2$.

\par 

Up until now, $\Omega$ was a free parameter of the forcing. One has been able to show that the lagrangian inertial-range spectrum does not depend upon $\Omega$ (only its extent does). Its value can be shown to depend upon the mean squared acceleration of the flow since, calculating \mbox{$\langle a^2\rangle=\int \omega^2 E_\mathrm{lag}(\omega) \mathrm{d}\omega$} from eq.(\ref{ee2}), it is found that $\Omega=\frac{\langle a^2\rangle}{\varepsilon}$. This is a clear indication of the entanglement between the flow and the stirring described earlier, where the lagrangian stirring force correlation time $\propto\Omega^{-1}$ depends upon the mean squared acceleration and dissipation which are physical values of the flow.


\begin{thebibliography}{10}

\bibitem{kolmo}
A.~{Kolmogorov}.
\newblock {The Local Structure of Turbulence in Incompressible Viscous Fluid
  for Very Large Reynolds' Numbers}.
\newblock {\em Akademiia Nauk SSSR Doklady}, 30:301--305, 1941.

\bibitem{kolmo62}
A.~N. Kolmogorov.
\newblock A refinement of previous hypotheses concerning the local structure of
  turbulence in a viscous incompressible fluid at high reynolds number.
\newblock {\em Journal of Fluid Mechanics}, 13(1):82, 1962.

\bibitem{obuk}
A.~M. {Obukhov}.
\newblock On the distribution of energy in the spectrum of turbulent flow
\newblock {\em Akademiia Nauk SSSR Doklady}, 32:19, 1941.

\bibitem{kraich1}
R.~H. Kraichnan.
\newblock The structure of isotropic turbulence at very high reynolds numbers.
\newblock {\em Journal of Fluid Mechanics}, 5(4):497?543, 1959.

\bibitem{kraich2}
R.~H. Kraichnan.
\newblock Lagrangian history closure approximation for turbulence.
\newblock {\em Phys. Fluids}, 8(4):575, 1965.

\bibitem{wyld}
H.~W. {Wyld Jr.}
\newblock {Formulation of the theory of turbulence in an incompressible fluid}.
\newblock {\em Ann. Phys.}, 14:143, 1961.

\bibitem{mccomb}
W.~D. McComb.
\newblock Theory of turbulence.
\newblock {\em Reports on Progress in Physics}, 58(10):1117, 1995.

\bibitem{yakhot}
Victor Yakhot and S.~A. Orszag.
\newblock Renormalization group analysis of turbulence. i. basic theory.
\newblock {\em Journal of Scientific Computing}, 1(1):3--51, 1986.

\bibitem{kaneda1}
Y.~Kaneda.
\newblock Renormalized expansions in the theory of turbulence with the use of
  the lagrangian position function.
\newblock {\em Journal of Fluid Mechanics}, 107:131, 1981.

\bibitem{kaneda2}
Y.~Kaneda.
\newblock Inertial range structure of turbulent velocity and scalar fields in a
  lagrangian renormalized approximation.
\newblock {\em The Physics of Fluids}, 29(3):701--708, 1986.

\bibitem{lvov}
V.~S. L'vov.
\newblock Scale invariant theory of fully developed hydrodynamic
  turbulence-hamiltonian approach.
\newblock {\em Physics Reports}, 207(1):1 -- 47, 1991.

\bibitem{instanton}
G.~Falkovich, I.~Kolokolov, V.~Lebedev, and A.~Migdal.
\newblock Instantons and intermittency.
\newblock {\em Phys. Rev. E}, 54:4896--4907, Nov 1996.

\bibitem{falko3}
G.~Falkovich and A.~Zamolodchikov.
\newblock Operator product expansion and multi-point correlations in turbulent
  energy cascades.
\newblock {\em Journal of Physics A: Mathematical and Theoretical},
  48(18):18FT02, 2015.

\bibitem{polya}
A.M. Polyakov.
\newblock The theory of turbulence in two dimensions.
\newblock {\em Nuclear Physics B}, 396(2):367 -- 385, 1993.

\bibitem{kraich70}
Robert~H. Kraichnan.
\newblock Diffusion by a random velocity field.
\newblock {\em The Physics of Fluids}, 13(1):22--31, 1970.

\bibitem{falko}
G.~Falkovich, Gawedsky K., and Vergassola M.
\newblock Particles and fields in fluid turbulence.
\newblock {\em Rev. Mod. Phys.}, 73:913, 2001.

\bibitem{falko2}
G.~Falkovich, I.~Fouxon, and Y.~Oz.
\newblock New relations for correlation functions in navier-stokes turbulence.
\newblock {\em Journal of Fluid Mechanics}, 644:465?472, 2010.

\bibitem{sato}
Yukinari Sato and Kazuo Yamamoto.
\newblock Lagrangian measurement of fluid-particle motion in an isotropic
  turbulent field.
\newblock {\em Journal of Fluid Mechanics}, 175:183\u2013199, 1987.

\bibitem{mord}
O.~Michel N.~Mordant, P.~Metz and J.-F. Pinton.
\newblock Measurement of lagrangian velocity in fully developed turbulence.
\newblock {\em Phys. Rev. Lett.}, 87:214501, 2001.

\bibitem{popesim}
P.~K. Yeung, S.~B. Pope, and B.~L. Sawford.
\newblock Reynolds number dependence of lagrangian statistics in large
  numerical simulations of isotropic turbulence.
\newblock {\em Journal of Turbulence}, 7:N58, 2006.

\bibitem{bif1}
L.~Biferale, E.~Bodenschatz, M.~Cencini, A.~S. Lanotte, N.~T. Ouellette,
  F.~Toschi, and H.~Xu.
\newblock Lagrangian structure functions in turbulence: A quantitative
  comparison between experiment and direct numerical simulation.
\newblock {\em Physics of Fluids}, 20(6):065103, 2008.

\bibitem{falk3}
Gregory Falkovich, Haitao Xu, Alain Pumir, Eberhard Bodenschatz, Luca Biferale,
  Guido Boffetta, Alessandra~S. Lanotte, Federico Toschi, and
  (International~Collaboration for Turbulence~Research).
\newblock On lagrangian single-particle statistics.
\newblock {\em Physics of Fluids}, 24(5):055102, 2012.

\bibitem{lagfd}
A.~Bennett.
\newblock Lagrangian fluid dynamics.
\newblock {\em Cambridge University Press}, page~63, 2006.

\bibitem{ghid}
B.~{Brefort}, J.~M. {Ghidaglia}, and R.~{Temam}.
\newblock Attractors for the penalized navier-stokes equations.
\newblock {\em SIAM Journal of Mathematical Analysis}, 19:1--21, January 1988.

\bibitem{qft}
M.~{Maggiore}.
\newblock A modern introduction to quantum field theory.
\newblock {\em Oxford University Press}, page 180, 2005.

\bibitem{alphau}
Alain Pumir Eberhard Bodenschatz Luca Biferale Guido Boffetta Alessandra S.
  Lanotte Federico~Toschi Gregory~Falkovich, Haitao~Xu.
\newblock On lagrangian single-particle statistics.
\newblock {\em Phys. Fluids}, 24:055102, 2012.

\end{thebibliography}
\end{document}